# Reversal of the lattice structure in SrCoO$_x$ epitaxial thin films studied by real-time optical spectroscopy and first-principles calculations

Woo Seok Choi,[1,†] Hyoungjeen Jeen,[1,†] Jun Hee Lee,[2] S. S. Ambrose Seo,[3]

Valentino R. Cooper,[1] Karin M. Rabe,[4] and Ho Nyung Lee[1,*]

[1]*Materials Science and Technology Division, Oak Ridge National Laboratory, Oak Ridge, TN 37831, USA*

[2]*Department of Chemistry, Frick Laboratory, Princeton University, Princeton, NJ 08544, USA*

[3]*Department of Physics and Astronomy, University of Kentucky, Lexington, KY 40506, USA*

[4]*Department of Physics and Astronomy, Rutgers University, Piscataway, NJ 08854, USA*

Using real-time spectroscopic ellipsometry, we directly observed a reversible lattice and electronic structure evolution in SrCoO$_x$ ($x = 2.5 - 3$) epitaxial thin films. Drastically different electronic ground states, which are extremely susceptible to the oxygen content $x$, are found in the two topotactic phases, i.e. the brownmillerite SrCoO$_{2.5}$ and the perovskite SrCoO$_3$. First principles calculations confirmed substantial differences in the electronic structure, including a metal-insulator transition, which originates from the modification in the Co valence states and crystallographic structures. More interestingly, the two phases can be reversibly controlled by changing the ambient pressure at greatly reduced temperatures. Our finding provides an important pathway to understanding the novel oxygen-content-dependent phase transition uniquely found in multivalent transition metal oxides.

Oxygen stoichiometry and valence state of the metal cations in transition metal oxides (TMOs) play an essential role in determining major aspects of their physical properties and behaviors, including color, spin states, and electronic structures [1,2]. The multivalent nature of most transition metals often causes the formation of various solid TMO phases, yielding rich and intriguing oxygen content-dependent phase diagrams [3-5]. The multivalency also facilitates the redox reaction in many TMOs making them attractive for many technological applications. More specifically, the oxygen content-dependent valence states and physical property transitions are closely related to the ionic conduction and catalytic activities, which are basic functions in most cutting-edge energy storage and generation devices [5-8]. Exploring the role of oxygen in energy-related TMOs would therefore provide invaluable insight in identifying new potential applications of these materials.

Among TMOs, $SrCoO_x$ (SCO, $x = 2.5 - 3$) is an excellent candidate for studying the role of oxygen content in determining its physical properties. In contrast to $LaCoO_3$, whose $Co^{3+}$ valence state is so robust that the formation of oxygen vacancies is rather difficult [9], SCO has oxygen content-dependent characteristic crystal structures, which result in drastic changes in its electronic and magnetic structures [4,10-12]. Moreover, while the prototypical and most widely used cobaltite $(La^{3+},Sr^{2+})(Co^{3+},Co^{4+})O_{3-\delta}$ has two different tuning knobs to control the valence state of Co, $i.e.$, A-site cation and oxygen, SCO has only $Sr^{2+}$ in the A-site. This provides an unambiguous route to addressing the link between the oxygen content and valence state change in TMOs.

SCO forms the brownmillerite (BM) structure for $x = 2.5$, as schematically shown in the top center inset of Fig. 1(a). BM-SCO has alternating $CoO_6$ octahedral and $CoO_4$ tetrahedral sub-layers stacked along the $c$-axis, forming 1D oxygen vacancy-like channels. The overall Co valence state is 3+ with $3d^6$ electrons, producing an electronically insulating ground state [4]. On the other hand, when $x = 3.0$, SCO stabilizes as a perovskite (P), as shown in the top left inset of Fig. 1(a) [12]. Now the Co valence state is 4+ with $3d^5$ electrons, resulting in a metallic ground state [12,13]. While it is difficult to grow a fully oxygenated P-SCO film or single crystal, theoretical calculations have suggested that it has rich strain-induced electronic and magnetic phases, including a ferroelectric one [14].

In this letter, we report an unprecedented approach to directly address and study the topotactic phase transformation between BM- and P-SCO during redox reactions. A topotactic phase change refers to the structural transformation within a crystal lattice, which may include loss or gain of element, so that the final phase has one or more crystallographically equivalent framework of the parent phase. The phase reversal between BM- and P-SCO was investigated using optical spectroscopy and first principles calculations. The optical properties and electronic structures of SCO were found to be highly sensitive to a change in the oxygen stoichiometry, exhibiting a metal-insulator transition (MIT).

We used pulsed laser epitaxy to grow SCO ($x$ = 2.5 and 3.0) thin films on (001) $(LaAlO_3)_{0.3}(SrAl_{0.5}Ta_{0.5}O_3)_{0.7}$ (LSAT) and (001) $SrTiO_3$ (STO) substrates. Note that the optical responses were qualitatively the same for the same phase films grown on different substrates, suggesting that the strain state does not play a significant role in determining the optical properties [9]. See Supplemental Material for more details on experimental approaches and theoretical calculations [15].

Figure 1(a) shows optical absorption spectra as a function of photon energy, $a(\omega)$, extracted from transmittance measurements at 300 and 5 K for both BM- and P-SCO epitaxial films. The overall spectral features at 300 K remain the same down to 5 K, indicating that the electronic structure of SCO is, at least at low temperatures, without any discernible optical phase transitions. BM-SCO shows an insulating $a(\omega)$ characteristic, and by plotting $a^{1/2}(\omega)$ as shown in the bottom right inset of Fig. 1(a), we could determine a direct band gap of ~0.35 eV at 300 K. The band gap further opens up to ~0.45 eV at 5 K, indicating a clear insulating behavior. The optical band gap of the BM-SCO has not been reported previously, and we note that this value is much smaller compared to other bulk BM oxides, such as $BaInO_{2.5}$ (2.7 eV) or $SrFeO_{2.5}$ (2.0 eV) [16,17]. The small band gap is highly advantageous for many technological applications, e.g., for cathodes in solid oxide fuel cells, as it could realize a mixed ionic and electronic conductor at a moderate temperature. More importantly, due to the small band gap, SCO can readily undergo an MIT with a change in the Co valence state (or oxidation state). As shown in Fig. 1(a), P-SCO is indeed found to be metallic, with the Drude absorption at low photon energies. (Note that, even though it is highly unlikely based on the clear metallic behavior seen from the *dc* transport measurement, we do not exclude a possibility that the absorption might decrease as the photon energy further decreases. This can be assured by further investigation at low photon energies with far-infrared ellipsometry.) As temperature is lowered, the Drude peak is slightly enhanced, indicating a clear metallic optical response. Note that such outstanding optical properties from P-SCO have not been previously explored due to the lack of single crystalline materials. To further confirm the metallic property of P-SCO, we have also performed temperature-dependent resistivity measurements on the SCO films, as shown in Fig. 1(d). Metallic and insulating behaviors are clearly observed for P-SCO and BM-SCO, respectively, as expected from the optical spectra.

Two different peak features could be identified in $a(\omega)$ for each SCO phase: For BM-SCO, the peaks marked as "$\alpha$" and "$\beta$" at ~1.4 and ~2.8 eV, respectively; and for P-SCO, the Drude peak and the peak "$\gamma$" at ~1.7 eV. Based on the characteristic absorption peak features, we could construct schematic band diagrams for the two SCO phases as shown in Fig. 1(b) and 1(c), respectively. For BM-SCO, both octahedral ($Co^o$) and tetrahedral ($Co^t$) crystal fields should be considered. Thus, we attribute the peak $\alpha$ to a *d-d* transition between these various Co 3*d* states, taking into account their energy levels. Moreover, the peak $\beta$ is attributed to a *p-d* transition from the O 2*p* state to the Co 3*d* state. For P-SCO, the situation is simpler than that of BM-SCO, as only the octahedral crystal field is considered. Here, the Fermi level ($E_F$) crosses the Co 3*d* state indicating a metallic ground state. The peak $\gamma$ can then be attributed to a *p-d* transition from the O 2*p* state to the Co 3*d* state. Overall, our results suggest drastically different optical properties between the two SCO phases, where the main difference is the oxygen content.

In order to clarify the schematic band diagram and obtain more quantitative information, we performed density functional theory (DFT) calculations to map out the electronic structure evolution, as shown in Fig. 2. The total density of states (DOS) for BM- (Fig. 2(a)) and P-SCO (Fig. 2(g)) predict insulating and metallic ground states, respectively, consistent with our experimental observations. The projected DOS for BM- (Fig. 2(b-f)) and P-SCO (Fig. 2(h-j)) provide detailed information for identifying the origin of the peaks shown in Fig. 1(a). For BM-SCO, the highly distorted structure complicates the Co DOS, but the majority of the states near the $E_F$ are from Co $3d$ states. In particular, the peak $\alpha$ can be attributed to a $d$-$d$ transition between octahedral Co $t_{2g}$ states, and the peak $\beta$ to a transition between O $2p$ and mostly octahedral Co $t_{2g}$ and $e_g$ states. Note that our data on BM-SCO are similar to previously reported DFT results [18,19]. For P-SCO, the Drude peak originates from the Co $t_{2g}$ state crossing the $E_F$, and the peak $\gamma$ can be attributed to a transition between O $2p$ and Co $t_{2g}$ states, as illustrated in Fig. 1(c).

The metallic (and also ferromagnetic) phase of P-SCO is attributed to the double exchange between the hybridized $d$-states in Co ions [13,20]. However, even with a small external perturbation, e.g. strain or oxygen stoichiometry, this ferromagnetic metallic phase is known to compete with the antiferromagnetic insulating phase [14]. When oxygen vacancies start to form in P-SCO, and when they eventually order along the orthorhombic [010] direction to form the BM phase, the antiferromagnetic insulating phase seems to be more favorable. Microscopically, the formation of tetrahedral layers and large distortions of the CoO$_4$ tetrahedra in BM-SCO might disrupt the double exchange, leading to an insulating state. The incorporation of oxygen vacancies further splits the $t_{2g}$ and $e_g$ bands. The split $t_{2g}$ band opens a gap, results in the peak $\alpha$, and also seems to push the charge transfer gap to a higher energy for peak $\beta$ in BM-SCO.

Based on the DOS results, we further calculated the imaginary part of dielectric permittivity ($\varepsilon_2(\omega)$), for P- and BM-SCO, as shown in Fig. 2(k). Note that $\varepsilon_2(\omega)$ reflects the dissipation (or loss) of energy within the medium and is related to $a(\omega)$ through the following relationships, $a(\omega) = 4\pi k/\lambda$ and $\hat{\varepsilon} = \varepsilon_1 + i\varepsilon_2 = \hat{n}^2 = (n + ik)^2$, where $n$, $k$, and $\lambda$ represents the refractive index, the extinction coefficient, and the wavelength, respectively. Therefore, $\varepsilon_2(\omega)$ qualitatively reproduces $a(\omega)$ with qualitatively similar peak features. Indeed, the theoretically calculated $\varepsilon_2(\omega)$ in Fig. 2(k) consistently shows clear features of the peak $\beta$ and $\gamma$ seen experimentally in Fig. 1(a), respectively, for the BM- and P-SCO films.

A direct comparison between experimental and theoretical $\varepsilon_2(\omega)$ was possible by obtaining photon energy dependent spectra of the real ($\varepsilon_1$) and imaginary ($\varepsilon_2$) parts of the dielectric function using spectroscopic ellipsometry at room temperature, as shown in Fig. 3. Note that $\varepsilon_1(\omega)$ and $\varepsilon_2(\omega)$ show typical spectral shapes predicted from the Kramers-Kronig relation [21], further validating our measurements and analyses. Although the details on the Drude feature from P-SCO could not be completely unveiled due to the experimental limit of the spectral range in ellipsometry, the peaks $\beta$ and $\gamma$ were clearly visible in $\varepsilon_2(\omega)$, showing a qualitative similarity (e.g., in overall shape and similar peak positions) with the theoretically predicted data shown in Fig. 2(k). These peak features are the optical signatures of the two different oxygen contents in the BM- and P-SCO epitaxial thin films.

Interestingly, the two distinctively different SCO phases could be obtained reversibly from one to the other by simple oxygen insertion into the BM-SCO or extraction out of the P-SCO thin films. The oxygen movement in the thin films naturally accompanies optical and electronic phase transition. This observation underscores an unconventional but important topotactic phase transformation involving redox reaction in a sample, which is unique for epitaxial TMO thin films which contain multivalent cations. The evolution of the electronic structure during reversible redox reactions in the material was directly monitored using a real-time spectroscopic ellipsometer, while simultaneously controlling the oxygen partial pressure ($P(O_2)$) and temperature in a vacuum chamber [22]. This technique enables us to study fundamental optical properties, during the redox activity and associated valence state changes in oxide thin films [22,23].

In order to elucidate the electronic structure evolution during the topotactic phase conversion, we first placed a P-SCO film in vacuum ($< 10^{-6}$ Torr) and raised the temperature to extract oxygen and transform the film into BM-SCO. Figure 4(a) depicts an $\varepsilon_2(\omega)$ map of the process as a function of temperature, clearly showing the phase transition: The peak $\gamma$ at ~1.5 eV from P-SCO starts to blue-shift at ~200 ℃ and disappears above ~300 ℃. Then the peak $\beta$ at 2.5 eV from BM-SCO starts to appear at 320 ℃ indicating the completion of the topotactic transformation from P to BM phase. This process suggests that oxygen in P-SCO becomes mobile at relatively low temperatures and can be pulled out of the thin film in vacuum. In addition, the time scale for the transformation is within tens of seconds, implying a very fast oxygen ion movement through the epitaxial thin films. It is quite intriguing that the oxygen movement in SCO can be so rapid even at relatively low temperatures. For example, in STO, a prototypical perovskite structure, a noticeable oxygen movement starts above 700 ℃ and hundreds of seconds are required to observe the resultant change in physical properties [24,25]. While the exact origin of this efficient oxygen movement is out of scope of the current letter, we note that the thermodynamic Gibbs free energy does not vary much for the formation of different SCO phases [26]. One possible explanation is that the unstable $Co^{4+}$ in P-SCO might easily weaken the Co-O bond strength within the $CoO_6$ octahedra, facilitating the free oxygen movement.

Once BM-SCO is formed, this phase is maintained robustly as long as the vacuum ambient is maintained. For example, it does not change back to P-SCO even after the temperature is further lowered as shown in Fig. 4(b). When the film is cooled to room temperature, only small temperature dependent spectral changes associated with the thermal broadening were noted without any significant phase transitions.

Conversely, we could also transform BM- into P-SCO by similarly raising the temperature in oxygen atmosphere (oxidizing condition). $P(O_2)$ of 500 Torr was used to convert BM- into P-SCO, as shown in Fig. 5(a). The peak $\beta$ for BM-SCO shows a red-shift and evolves into the peak $\gamma$ at a slightly higher temperature, compared to the transformation of P- into BM-SCO. Note that the transformation temperature should depend on various parameters, such as ambient pressure. Nevertheless, it is clear that a similar low temperature is required to activate or mobilize oxygen within SCO thin films, regardless of its original structure. Once a P-SCO is obtained from BM-SCO at high $P(O_2)$, it also stays robustly as P-SCO, as we lowered the temperature at the same $P(O_2)$. Furthermore, as shown in Fig. 5(b), we could successfully recover the BM-SCO phase, by re-annealing the sample in vacuum. Similar to what was observed in Fig. 4(a), we could recover the BM SCO at ~300 ℃, indicating that the two distinctively different phases can be reversibly cycled back and forth through oxygen insertion and extraction.

In conclusion, the correlation among the topotactic oxygen stoichiometry change, crystal structure, valence state, and electronic structure transition in Co-based TMO epitaxial thin films was explored. The direct observation of reversible electronic structure and valence state changes, including an MIT, provides unambiguous understanding on the evolution of a redox reaction in multivalent SCO epitaxial thin films. In particular, the reduced temperature and prompt topotactic phase conversion through the redox activity make SCO a promising candidate for a wide range of applications in electrochemical materials and devices.

We thank D. Shin for helpful discussions. This work was supported by the U.S. Department of Energy, Basic Energy Sciences, Materials Sciences and Engineering Division (W.S.C., H.J., V.R.C., and H.N.L.). This research used resources of the National Energy Research Scientific Computing Center, which is supported by the Office of Science of the U.S. Department of Energy under Contract No. DE-AC02-05CH11231 (V.R.C.). The work at the University of Kentucky was supported by the NSF through Grant EPS-0814194, the Center for Advanced Materials, and the Kentucky Science and Engineering Foundation with the Kentucky Science and Technology Corporation through Grant Agreement No. KSEF-148-502-12-303 (S.S.A.S.). [†]W.S.C. and H.J. contributed equally to this work.

*hnlee@ornl.gov

Figure captions

FIG. 1 (color online).  (a) $a(\omega)$ obtained from transmittance measurement for the BM and P-SCO single crystalline films. While BM-SCO shows insulating behavior, P-SCO shows metallic optical spectrum across the temperature region we have studied. Distinct peak features are denoted as $\alpha$ and $\beta$ for BM-SCO and $\gamma$ for P-SCO. The top insets show schematic crystal structures, where empty circles denote oxygen vacancy sites. The lower inset shows $a^{1/2}(\omega)$ spectra to obtain direct optical band gaps for BM-SCO. Schematic band diagrams of (b) BM and (c) P-SCO. For BM-SCO, $3d$ orbital states under tetragonal and octahedral crystal field are denoted as superscripts $t$ and $o$, respectively. (d) Typical temperature dependent resistivity behavior shown for the BM- and P-SCO thin films.

FIG. 2 (color online). (a) Total and (b-f), projected DOS onto (b) O2$p$, (c) Co $t_{2g}^{o}$, (d) Co $e_{g}^{o}$, (e) Co $t_{2g}^{t}$, and (f) Co $e_{g}^{t}$ for BM-SCO. (g) Total and (h-j) projected DOS onto (h) O2$p$, (i) Co $t_{2g}^{o}$, and (j) Co $e_{g}^{o}$ for P-SCO. (k) Calculated $\varepsilon_2(\omega)$ for BM (red) and P (blue) SCO phases.

FIG. 3 (color online). Experimental $\varepsilon_1(\omega)$ (dotted lines) and $\varepsilon_2(\omega)$ (solid lines) of SCO thin films, obtained from spectroscopic ellipsometry at room temperature. Peak $\beta$ for BM-SCO (red lines) and $\gamma$ for P-SCO (blue lines) corresponding to the optical absorption spectra in Fig. 1 are clearly seen in $\varepsilon_2(\omega)$.

FIG. 4 (color online). Temperature-dependent spectral evolution of P-SCO into BM-SCO. (a) $\varepsilon_2(\omega)$ maps of P-SCO in the process of annealing in vacuum to convert to BM-SCO, followed by (b) cooling in the same ambient.

FIG. 5 (color online). Reversible phase transformation. (a) $\varepsilon_2(\omega)$ map of BM-SCO in the process of annealing in 500 Torr of $P(O_2)$ to convert to P-SCO. After the conversion to the latter phase, the sample was cooled to room temperature. (b) $\varepsilon_2(\omega)$ map of converted P-SCO from (a) during re-annealing in vacuum to convert back to BM-SCO.

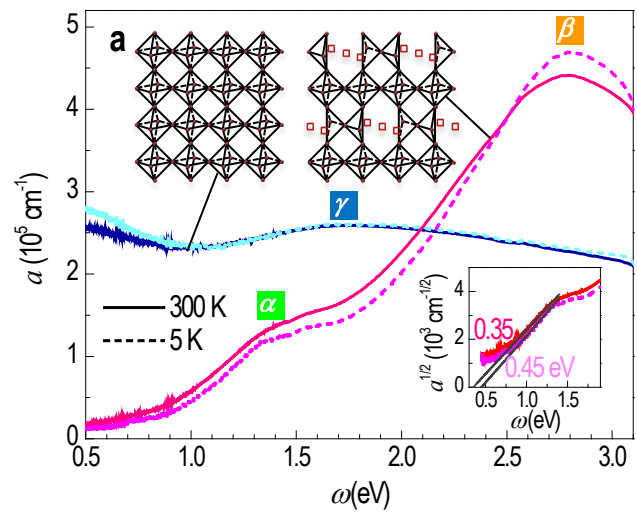
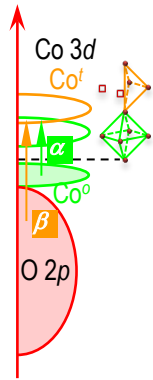
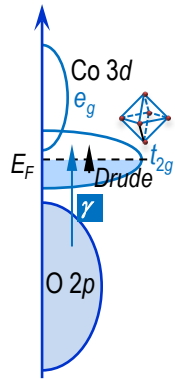
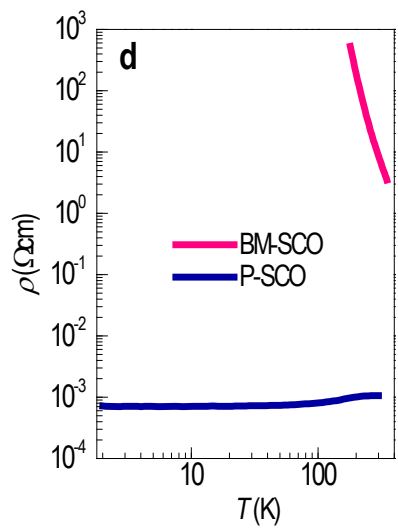

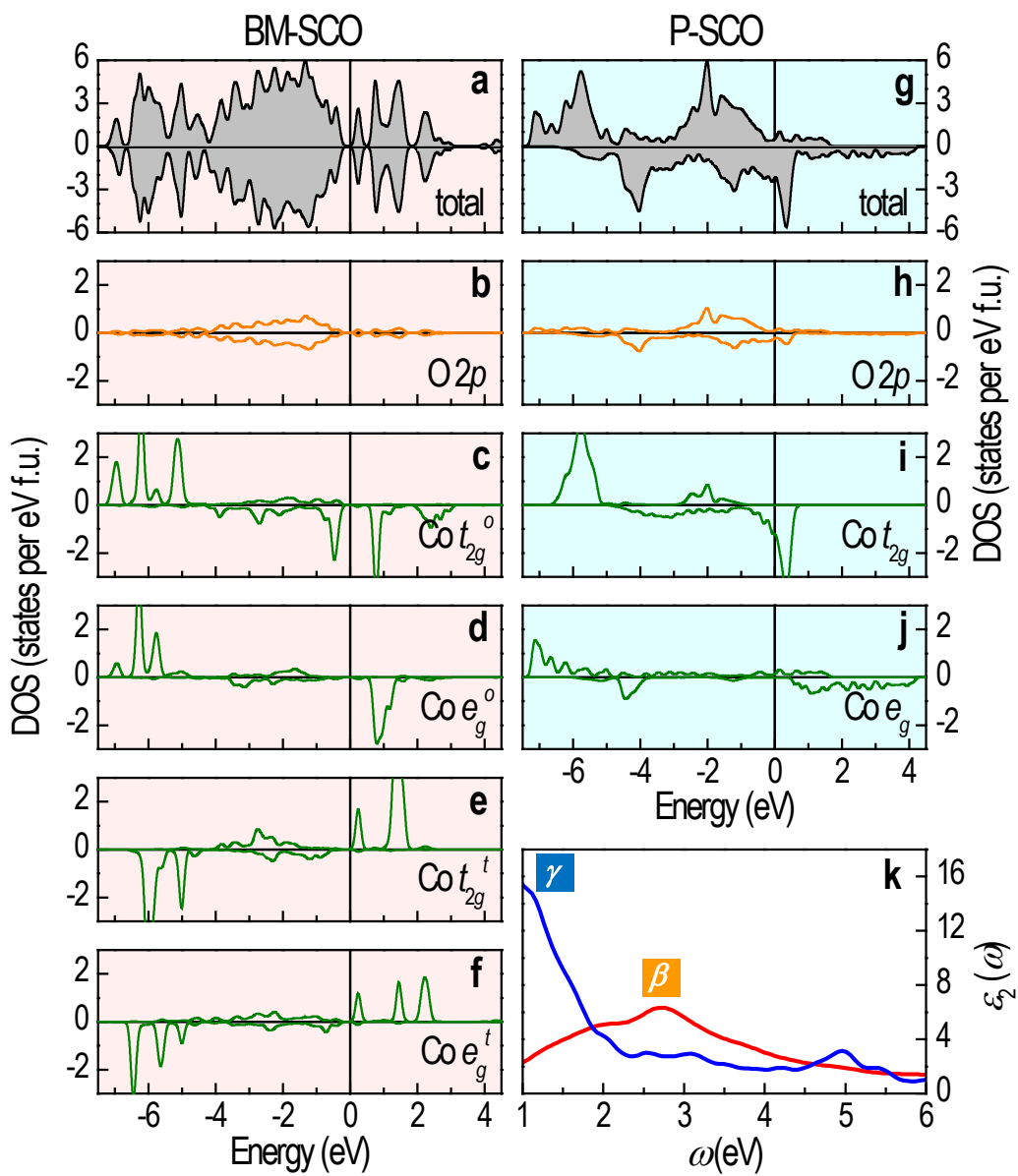

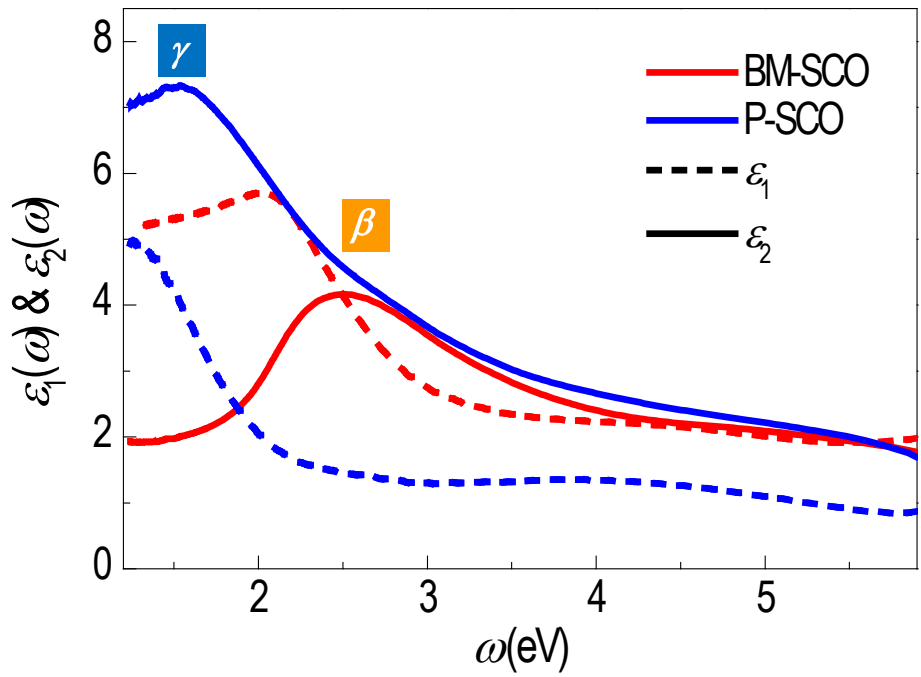

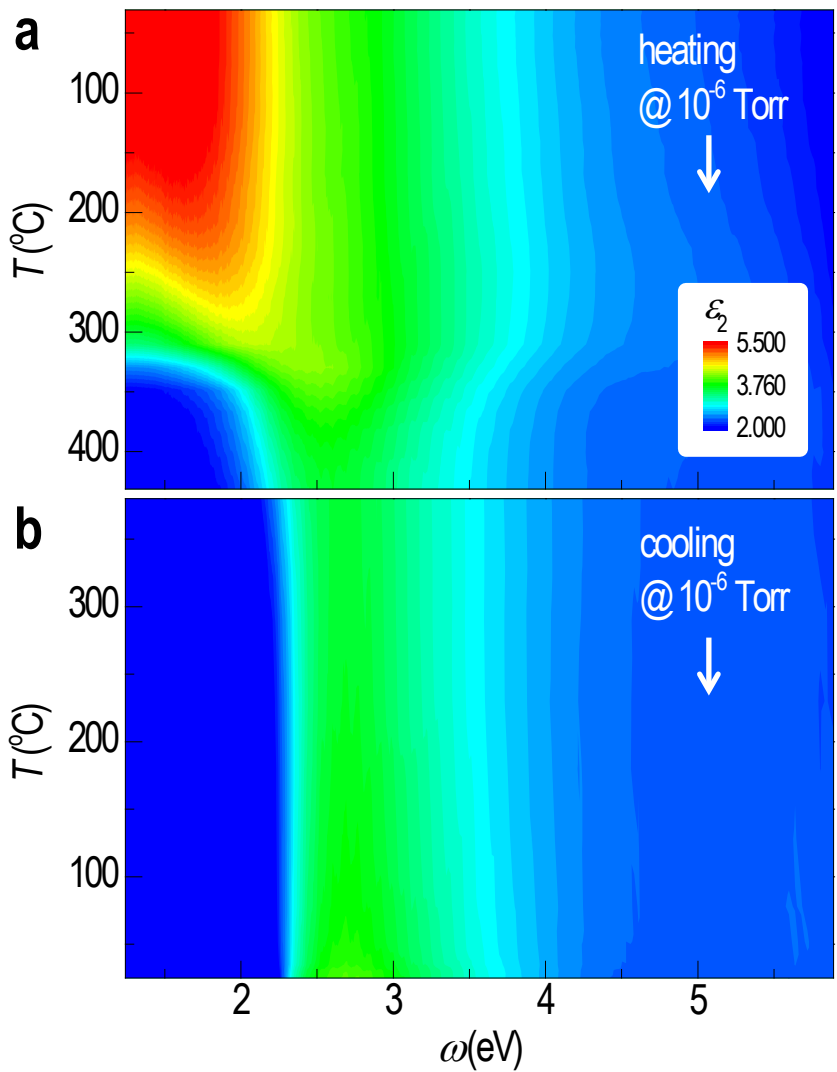

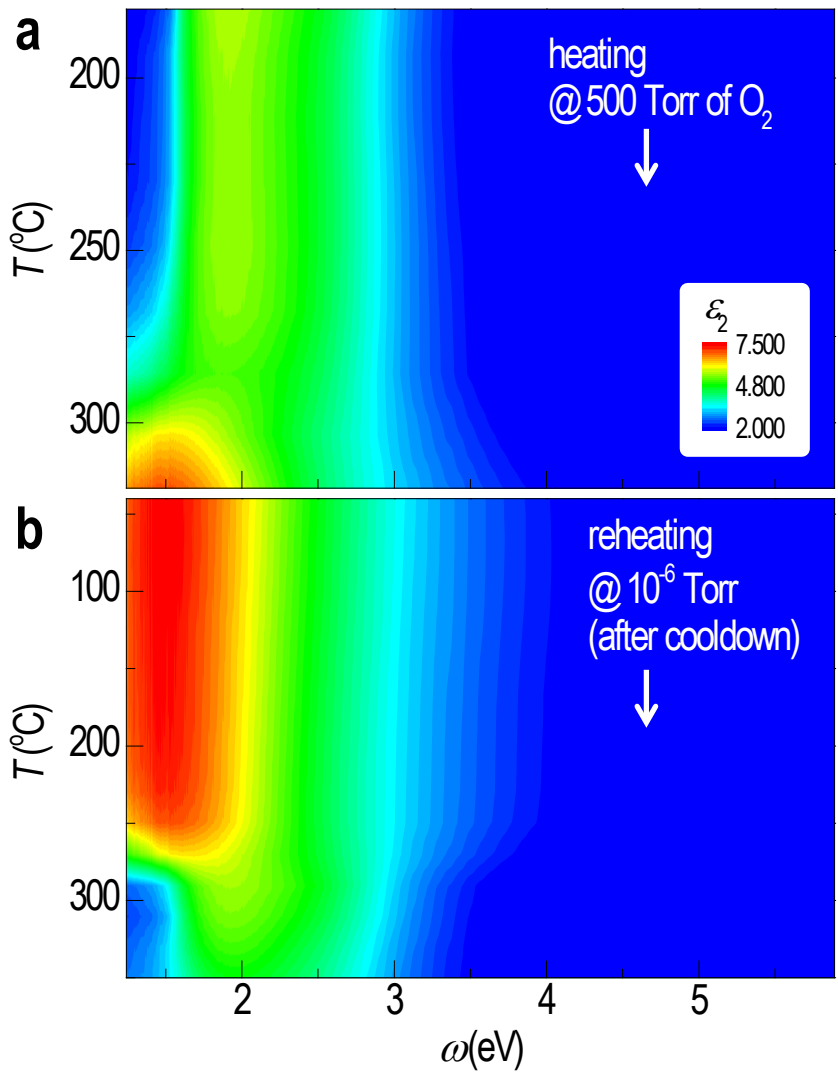

# Reversal of the lattice structure in SrCoO$_x$ epitaxial thin films studied by real-time optical spectroscopy and first-principles calculations

Woo Seok Choi,[1,†] Hyoungjeen Jeen,[1,†] Jun Hee Lee,[2] S. S. Ambrose Seo,[3]

Valentino R. Cooper,[1] Karin M. Rabe,[4] and Ho Nyung Lee[1,]*

[1]*Materials Science and Technology Division, Oak Ridge National Laboratory, Oak Ridge, TN 37831,*

*USA*

[2]*Department of Chemistry, Frick Laboratory, Princeton University, Princeton, NJ 08544, USA*

[3]*Department of Physics and Astronomy, University of Kentucky, Lexington, KY 40506, USA*

[4]*Department of Physics and Astronomy, Rutgers University, Piscataway, NJ 08854, USA*

## Details on Experimental Approaches and Theoretical Calculations

Temperature dependent resistivity curves, $\rho(T)$, were measured with the standard van der Pauw method, using a physical property measurement system (PPMS, Quantum Design Inc.). Optical transmittance measurements were performed for samples grown on double-side-polished STO substrates using a grating-type spectrophotometer (LAMBDA 950, PerkinElmer) between 0.4 and 3.1 eV. Low temperature spectra were taken using a continuous liquid-He flow optical cryostat (OptistatCF, Oxford) mounted on the spectrophotometer. Temperature and ambient controlled real-time spectroscopic ellipsometry was performed for samples grown on LSAT, using an ellipsometer (M-2000, J.A. Woollam

Co., Inc.) between 1.25 and 5.90 eV at an incident angle of ~66.5°, attached to a vacuum chamber with a lamp heater [S1]. The temperature ramping rate was ~0.1 °C/s, and each spectrum was measured within 2 seconds. We have also tested ellipsometry for multiple angles (50°, 60°, and 70°) at room temperature to confirm that the results from the real-time ellipsometry are valid. A simple two-layer model (film/substrate) was used to extract dielectric functions from the ellipsometric angles $\Psi$ and $\Delta$. Other detailed variations, such as intermixing layer at the interface or surface roughness layer caused minimal differences in the resultant dielectric function. Indeed, the surface roughness of the films was less than 1 nm as observed from atomic force microscopy. Finally, first-principles calculations were performed using density-functional theory within the generalized gradient approximation GGA + $U$ method with the Perdew-Becke-Erzenhof parameterization as implemented in the Vienna ab initio Simulation Package (VASP-5.2) [S2]. The projector augmentation wave (PAW) potentials [S3] include ten valence electrons for Sr ($4s^2 4p^6 5s^2$), nine for Co ($3d^8 4s^1$), and six for oxygen ($2s^2 2p^4$). The wave functions are expanded in a plane wave basis with 500 eV energy cutoff. For BM SCO, a $3 \times 1 \times 3$ Monkhorst-Pack $k$-point grid was used for relaxation and $6 \times 2 \times 6$ grid was used for optical property and density of states. For P-SCO, a $14 \times 14 \times 14$ grid was used. We use the Dudarev implementation [S4] with on-site Coulomb interaction $U = 4.5$ eV and on-site exchange interaction $J_{\mathrm{H}} = 1.0$ eV to treat the localized $d$ electron states in Co.